\documentclass[%
 reprint,
 amsmath,amssymb,
 aps,
]{revtex4-1}
\usepackage{float}
\usepackage{graphicx}
\usepackage{dcolumn}
\usepackage{bm}
\usepackage{color}
\usepackage{braket}
\usepackage{amsmath,amssymb}
\usepackage{subfig}
\usepackage{textcomp}
\begin{document}

\preprint{APS/123-QED}

\title{Laser Cooling of a Yb Doped Silica Fiber by 18 Kelvin From Room Temperature}

\author{Brian Topper$^{1,2}$}
\author{Mostafa Peysokhan$^{1,2}$}
\author{Alexander R. Albrecht$^{1}$}
\author{Angel S. Flores$^{3}$}
\author{Stefan Kuhn$^{4}$}
\author{Denny H{\"a}{\ss}ner$^{4}$}
\author{Sigrun Hein$^{4}$}
\author{Christian Hupel$^{4}$}
\author{Johannes Nold$^{4}$}
\author{Nicoletta Haarlammert$^{4}$}
\author{Thomas Schreiber$^{4}$}
\author{Mansoor Sheik-Bahae$^{1}$}
\author{Arash Mafi$^{1,2,*}$}
\email{mafi@unm.edu}
\affiliation{
$^1$University of New Mexico, Physics \& Astronomy and Interdisciplinary Science, 210 Yale Blv NE, Albuquerque, NM 87131, USA.\\
$^2$Center for High Technology Materials, University of New Mexico, 1313 Goddard St SE, Albuquerque, NM 87106, USA.\\
$^3$Air Force Research Laboratory, Directed Energy Directorate, 3550 Aberdeen Ave. SE, Kirtland Air Force Base, New Mexico 87117, USA.\\
$^4$Fraunhofer Institute for Applied Optics and Precision Engineering, Albert-Einstein-Str. 7, 07745 Jena, Germany.
}

\date{\today}

\begin{abstract}
A ytterbium doped silica optical fiber has been cooled by 18.4\,K below ambient temperature by pumping with 20\,W of 1035\,nm light in vacuum. In air, cooling by 3.6\,K below ambient was observed with the same 20\,W pump. The temperatures were measured with a thermal imaging camera and differential luminescence thermometry. The cooling efficiency is calculated to be $1.2\pm 0.1 \%$. The core of the fiber was codoped with Al$^{3+}$ for an Al to Yb ratio of 6:1, to allow for a larger Yb concentration and enhanced laser cooling. 
\end{abstract}

\maketitle

In the late 1920s, C. V. Raman discovered that when a material is exposed to light, its molecules scatter a small fraction of the incident photons inelastically. This inelastic scattering results in lower energy (Stokes) and higher energy (anti-Stokes) photons~\cite{raman1928}. Shortly after, Pringsheim postulated that anti-Stokes fluorescence may be used to decrease the temperature of a material ~\cite{pringsheim1929zwei}. It was not until the end of the 20\textsuperscript{th} century that optical cooling of solids was realized experimentally by Epstein and coworkers in ytterbium doped fluoride glass ~\cite{epstein1995observation}. Since this milestone achievement, systematic investigations have resulted in the observation of laser cooling in several families of rare-earth doped crystals and glasses~\cite{seletskiy2010laser, seletskiy2016laser,nemova2010laser, hoyt2000observation}. To date, the coldest temperature achieved by solid state optical refrigeration is in crystalline Yb:YLiF$_4$ down to 91\,K~\cite{melgaard2016solid}. For the first 24 years of laser cooling research activity, the observations of optically cooling glasses were confined to non-silicates ~\cite{seletskiy2016laser}. The paradigm has shifted recently with the success of cooling Yb-doped silica fibers and fiber preforms ~\cite{mobini2019laser,mobini2020laser,Knall:2020,10.1117/12.2510889,PhysRevApplied.11.014066,8426483,10.1117/12.2545233,10.1117/12.2548506,Knall:20,Knall_20_comp, Peysokhan:2021}. 

The high degree of polymerization and strong Si--O bonds make vitreous silica superior to fluoride systems, such as the ZLBAN family, with respect to mechanical and chemical durability. These attributes make silicates a more desirable material for fiber laser applications. In high-power fiber lasers, heat mitigation is required to maintain the integrity of the material and the beam profile~\cite{richardson2010review, brown2001,zenteno1993,ward2012,dawson2008, peysokhan2020characterization,peysokhan2019measuring}. Anti-Stokes florescence has been suggested as a viable method for heat mitigation in lasers~\cite{bowman1999, bowman2010, bowman2016}. Such a radiation-balanced fiber laser (RBL) experiences no increase in temperature, by effectively radiating out the waste heat generated during operation. Although silica-based radiation-balanced devices have been  reported this year in pioneering work ~\cite{knall2021radiationbalanced, knallRBL}, those devices are operating at orders of magnitude below the threshold of interest for adoption by industry.

Our work here demonstrates that it is possible for silica optical fibers to reach a steady-state of net cooling when exposed to pump powers of genuine interest to fiber laser practitioners. This suggests we are rapidly approaching the realization of a technologically desirable RBL. Here, we present to the best of our knowledge, a new record in the cooling of Yb-doped silica in vacuum by more than 18\,K from ambient temperature. Further, we observe record cooling in air by more than 6\,K from ambient, which is two orders of magnitude greater than previously published cooling results of optical fibers in air. We achieve this by using pump powers in the range of 1\,W to 185\,W at 1035\,nm wavelength.

The high-purity fibers (Table \ref{tab:material}) were drawn from preforms fabricated with the modified chemical vapor deposition technique. Cation (Yb, Al) doping of the core was carried out by the gas-phase doping technique ~\cite{kuhn2019}, using Yb(thd)$_3$ and AlCl$_3$ as precursors. Relative to previously successful laser cooling compositions doped with Al and F ~\cite{mobini2020laser}, the molar concentration of Yb$_2$O$_3$ was increased by 25\% for fiber A. These glasses were developed for single-mode, high-power fiber laser applications, thus a controlled core-cladding refractive index step is essential. To achieve this, codoping with fluorine was used to decrease the refractive index of the material and assure single mode operation in the drawn large-mode area double clad fiber geometry (e.g. 20/400 geometry).
\begin{table}
\centering
\caption{\bf Material properties of Yb doped fibers}
\begin{tabular}{ccc}
\hline
\hline
Fiber & A & B \\
\hline
Codopants & Al, F & Al, F \\
Yb$_2$O$_3$ (mol\%) & 0.15 & 0.12 \\
Yb$^{3+}$ density (10$^{25}$ atoms/m$^{3}$) & 6.56  & 5.26 \\
Al:Yb ratio & 6:1 & 8.3:1\\
NA$_{\rm core}$ & 0.06 & 0.05 \\
D$_{\rm core}$/D$_{\rm cladding}$
($\mu$m/$\mu$m)  & 900/1000 & 900/1000 \\
\hline
\hline
\end{tabular}
  \label{tab:material}
\end{table}

Fiber lasers using these types of glasses have been used to achieve continuous wave (CW) output powers of more than 4 kW from a single fiber, while maintaining good beam quality ~\cite{beier2017, beier2018}. Output powers like these can only be accomplished with, among other things, high-purity core materials with low background absorption. The background losses of these glasses were below 10\,dB\,km$^{-1}$ measured at a wavelength of 1200\,nm, which has been found to be acceptable for laser cooling silica ~\cite{mobini2020laser, Peysokhan:2021}.

The fabricated preforms consisted of a >3\,mm diameter doped core and >14\,mm undoped cladding. Partial removal of the cladding has enabled greater cooling in the study of a previous preform ~\cite{Peysokhan:2021}. Here, in effort to increase the cooling effect in these fibers, the undoped cladding was reduced from the preform and only a thin layer of passive cladding was left surrounding the active core material. Afterward, the preforms were drawn to fibers with an outer diameter of 1000\,$\mu$m and a doped core diameter of 900\,$\mu$m.

\begin{figure}[htbp]
\centering
 \includegraphics[width=3.25 in]{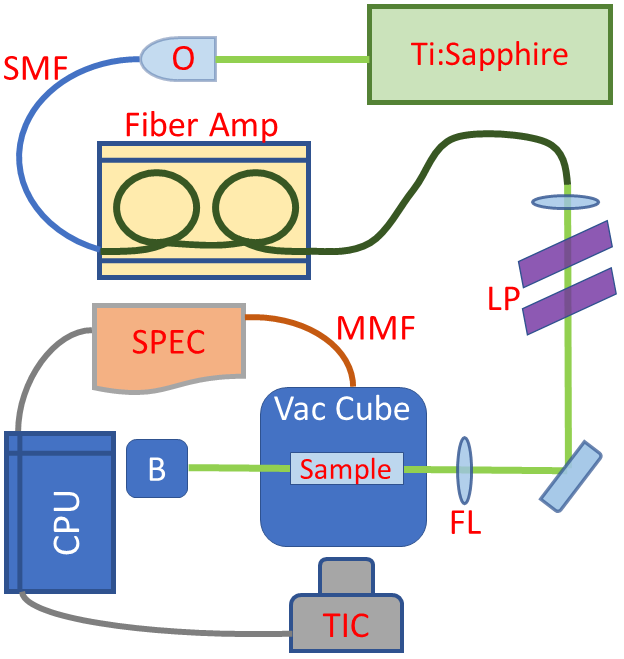}
\caption{Schematic of experimental set up with 1035 nm source (Ti:Sapphire), 20x objective lens (O), single mode fiber (SMF), custom-built fiber amplifier (Fiber Amp), two long pass filters (LP), focusing lens (FL), vacuum chamber (Vac Cube) containing the sample and thermally transparent windows for observation, thermal imaging camera (TIC), multimode fiber (MMF) connected to a spectrometer (SPEC), beam block (B), and computer for data collection (CPU). }
\label{fig:exp}
\end{figure}

For fiber A, the details of the cooling experiment are akin to those used in Ref.~\cite{Peysokhan:2021} and will only be briefly summarized here. Approximately 45\,mW of 1035\,nm light from a CW\,Ti:Sapphire laser is coupled through free space to a single mode fiber with an objective lens. A custom-built fiber amplifier increases the signal power, providing an output adjustable between 1\,W and 20\,W. The amplified signal serves as the input power to the Yb doped silica fiber subject to cooling. The doped fiber is supported by thin fused silica fibers to minimize conductive heating. To minimize convective heating, the doped fiber and sample holder are placed in an evacuated chamber where, for these experiments, a pressure of about 5\,$\times$\,10$^{-5}$\,torr was achieved. The input is coupled to the enclosed fiber by a 10\,cm focal length lens. Perpendicular to the axis of the fiber are two windows for real-time observation. One window is thermally transparent KCl for recording the temporal behavior of the fiber with a thermal imaging camera (TIC). The other window is fused silica through which discrete measurements of the fluorescence are taken by collecting the emitted light with a multimode fiber connected to a silicon CCD line spectrometer (Ocean Optics). See Fig.~\ref{fig:exp} for a visualization. For fiber B, an independent set of measurements were obtained using an amplifier capable of reaching 185\,W output of 1033\,nm light. The measurements on fiber B were made in air and data acquisition was carried out with a FLIR T540 thermal camera. 

\begin{figure}[htbp]
\centering
 \includegraphics[width=3.25 in]{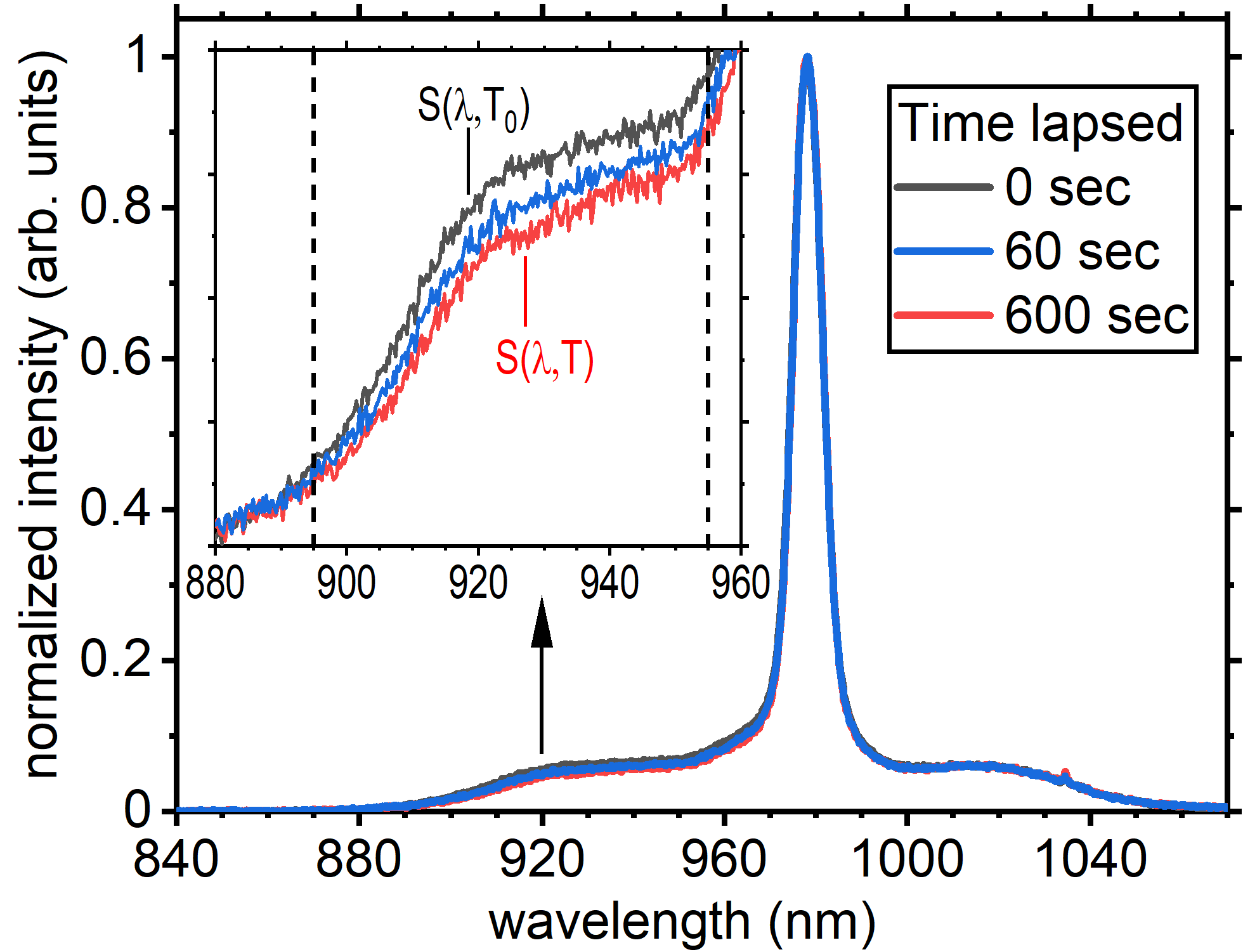}
\caption{ Yb:Silica emission spectra changing with temperature for fiber A. The inset is over the range 880\,nm and 960\,nm to highlight the domain relevant to determining the temperature of this fiber, indicated by the vertical dashed lines. The spectra S($\lambda$,T) are individually normalized to their intensity at $\lambda=978\,{\rm nm}$  and compared to the normalized spectrum at the start ($t=0$) of the experiment S($\lambda$,T$_0$) after exciting with 20\,W of coherent 1035\,nm light.}
\label{fig:spec}
\end{figure}

\begin{figure}[t]
\centering
 \includegraphics[width=3.25 in]{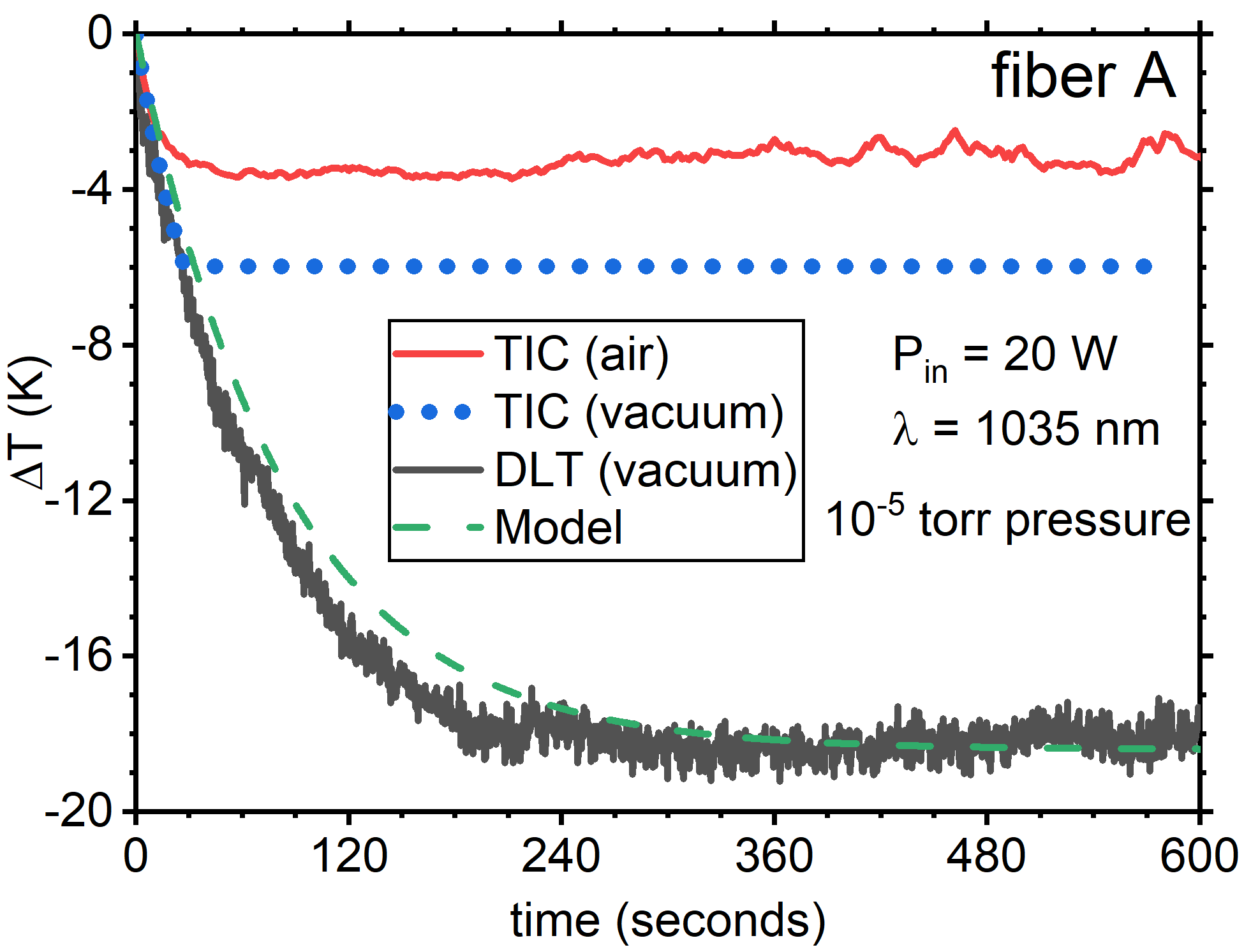}
\caption{Comparison of temperature measurements of fiber A pumped with 20\,W of 1035\,nm light in both vacuum (TIC-blue and DLT-black) and atmospheric (TIC-red) pressure conditions. The green line corresponds to the application of Eq. 3 to the DLT in vacuum data.}
\label{fig:cool}
\end{figure}

First, we discuss the cooling of fiber A. The thermal imaging camera was used to track the evolution of the temperature as was done in Refs.\,\cite{mobini2020laser}\,{\&}\,\cite{Peysokhan:2021}. The temperature difference is defined as $\Delta T=T-T_0$ where $T_0$ is taken to be the ambient temperature of 296\,K. To record cooling beyond $\Delta T=10\,{\rm K}$, differential luminescence thermometry (DLT) was employed~\cite{imangholi2006,Peysokhan:2021}. DLT exploits the temperature-dependence of the luminescence spectral form (see Fig.\,\ref{fig:spec}) which is dictated by the density of states.  
In the DLT analysis, each spectrum is normalized to its maximum (at $\lambda=978\,{\rm nm}$) to eliminate influence of input power fluctuations. The difference between a spectrum at time $t$ is then taken with respect to the spectrum taken at the onset of the experiment, where the cooling is assumed to be negligible and thus the spectral density is representative of $S(\lambda,T_0=296\,{\rm K}$). The normalized difference spectra is defined as
\begin{align}
\label{Eq:spectraldensity-1}
\Delta S(\lambda,T,T_0) = \frac{S(\lambda,T)}{S_{max}(T)} - \frac{S(\lambda,T_0)}{S_{max}(T_0)}.
\end{align}
The change in temperature has been found to be linearly proportional to the integrated difference in spectral density given by
\begin{align}
\label{Eq:spectralintegral-2}
S_{DLT}(T,T_0) =  \int_{\lambda_1}^{\lambda_2} \lvert \Delta S(\lambda,T,T_0) \rvert d\lambda ,
\end{align}
such that $\Delta {\rm T}= \alpha\, S_{DLT}$ with $\alpha=34.5\pm 0.4$\,K for fiber A. The temperature difference measured by the TIC is compared to the temperature difference by DLT in Fig.~\ref{fig:cool} for a 20\,W pump power incident on the fiber held under vacuum. In Fig.~\ref{fig:cool}, we see that, in the absence of convective heating contributions, both DLT (black line) and TIC (blue circles) measurements record nearly the same rate of cooling up to ~25 seconds lapsed time. At this point, the thermal imaging camera becomes saturated. When the experiment was performed in-air (red line), the maximum temperature difference achieved was less than 4\,K and so the TIC was able to reliably track the cooling of the fiber with 20\,W of input power under atmospheric pressure. Despite the maximum $\Delta T$ being reduced by about a factor of 5 for the in-air measurement, the onset of cooling for in-air and in-vacuum trials coincides for the initial ${\sim}$5\,seconds, $\Delta T {\approx} 1.3\,{\rm K}$, before the air measurement deviates. The dashed green line in Fig.~\ref{fig:cool} represents the application of the following exponential form to the data,
\begin{align}
\label{Eq:tempform-3}
&\Delta T(t) = \Delta T_{max}(e^{-t/\tau_c}-1).
\end{align}

$\Delta T_{max}$ is obtained from the experimental data over a continuous time interval greater than or equal to two minutes in duration while steady-state behavior is exhibited. The time constant in the exponential of  Eq.\,(\ref{Eq:tempform-3}) may be defined by 
\begin{align}
\label{Eq:terms-4}
 \tau_c=\frac{\rho V c_v}{4\epsilon \sigma T_0^3 A },
\end{align}
where $A$ is the area of the sample, $c_v=741\,{\rm J\cdot kg^{-1}\cdot K}^{-1}$ is the specific heat of fused silica, $\epsilon=0.85$ is the emissivity of the doped glass, $\rho=2.2\times 10^3\,{\rm kg\cdot m^{-3}}$ is the density of fused silica, $\sigma = 5.67\times 10^{-8}\,{\rm W\cdot m^{-2}\cdot K^{-4}}$ is the Stefan-Boltzmann constant, $T_0$ is the ambient temperature, and $V$ is the volume of the fiber. For the given fiber geometry, evaluation of Eq.\,(\ref{Eq:terms-4}) gives $\tau_c=81$\,s. This agrees well with the average experimental value $\tau_c = 84 {\pm} 3$\,s. Taking $\tau_c = 84 {\pm} 3$\,s and determining the $\Delta T_{max}$ from the TIC or DLT data, Eq.\,(\ref{Eq:tempform-3}) was found to model the experimental data quite well. We next determine the absorbed power, $P_{abs}$, with the Beer-Lambert law
\begin{align}
\label{Eq:spectraldensity-1}
P_{abs} = P_{in} \textsf{T$_{tot}$}
(1-e^{-\alpha_r l}).
\end{align}
$P_{in}$ is the pump power measured before the focusing lens, \textsf{T$_{tot}$} is the total transmission coefficient, $l$ is the length of the fiber, and $\alpha_r$ is the resonant absorption coefficient. \textsf{T$_{tot}$} is the product of the transmission of the focusing lens (\textsf{T$_{\rm l}$} = 0.998), the transmission of the chamber window (\textsf{T$_{\rm cw}$}=0.92), and the transmission into the glass fiber after accounting for Fresnel losses at the surface (\textsf{T$_{\rm g}$} = 0.96) such that \textsf{T$_{tot}$} = \textsf{T$_{\rm l}$}\textsf{T$_{\rm cw}$}\textsf{T$_{\rm g}$}. The resonant absorption coefficient was found to be $\alpha_r$($\lambda$=1035\,nm) = 1.92 $\pm$ 0.04 m$^{-1}$. The magnitude of the cooling of fiber A in-vacuum was found to increase with increasing absorbed power. With the absorbed power now known for each trial, we next inspect the slope of the $\Delta T(t)$ curves at $t=0$ to find the cooling efficiency, $\eta_c$, of fiber A at 1035\,nm wavelength via

\begin{align}
\label{Eq:coolingeff}
&\eta_c = \frac{- \rho V c_v}{P_{abs}} \partial_t \Delta T \rvert_{t=0}.
\end{align}

\begin{figure}[t]
\centering
 \includegraphics[width=3.25 in]{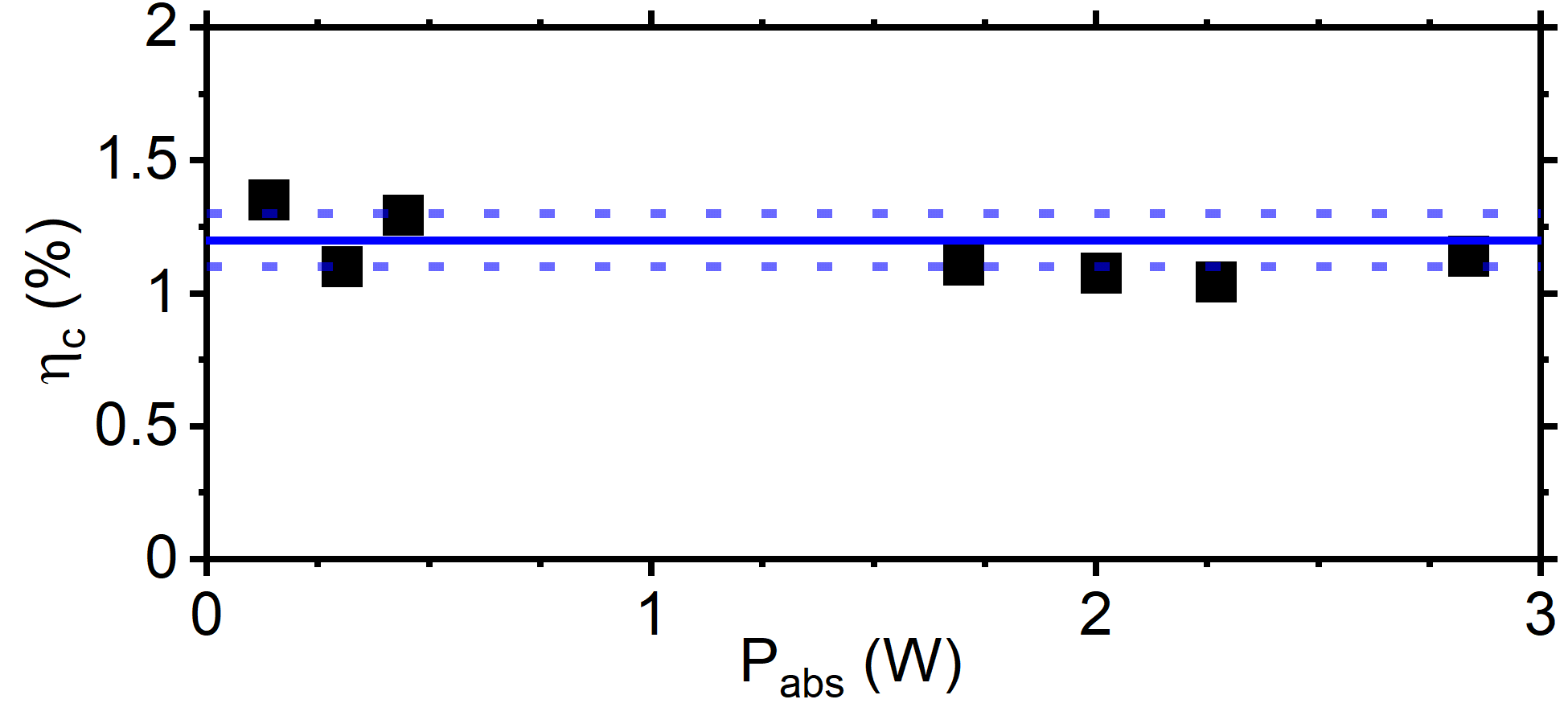}
\caption{Calculated cooling efficiency (fiber A) correlation with increasing absorbed power alongside the mean (solid blue line) and the standard deviation (blue dotted line).}
\label{fig:etac}
\end{figure}

The TIC data was used to calculate $\eta_c$ for each trial, as the $\Delta T(t)$ for small $t$ was below the saturation limit. We find the cooling efficiency of fiber A to be $\eta_c = 1.2 \pm 0.1 \%$ (Fig. \ref{fig:etac}).

\begin{figure}[t]
\centering
 \includegraphics[width=3.25 in]{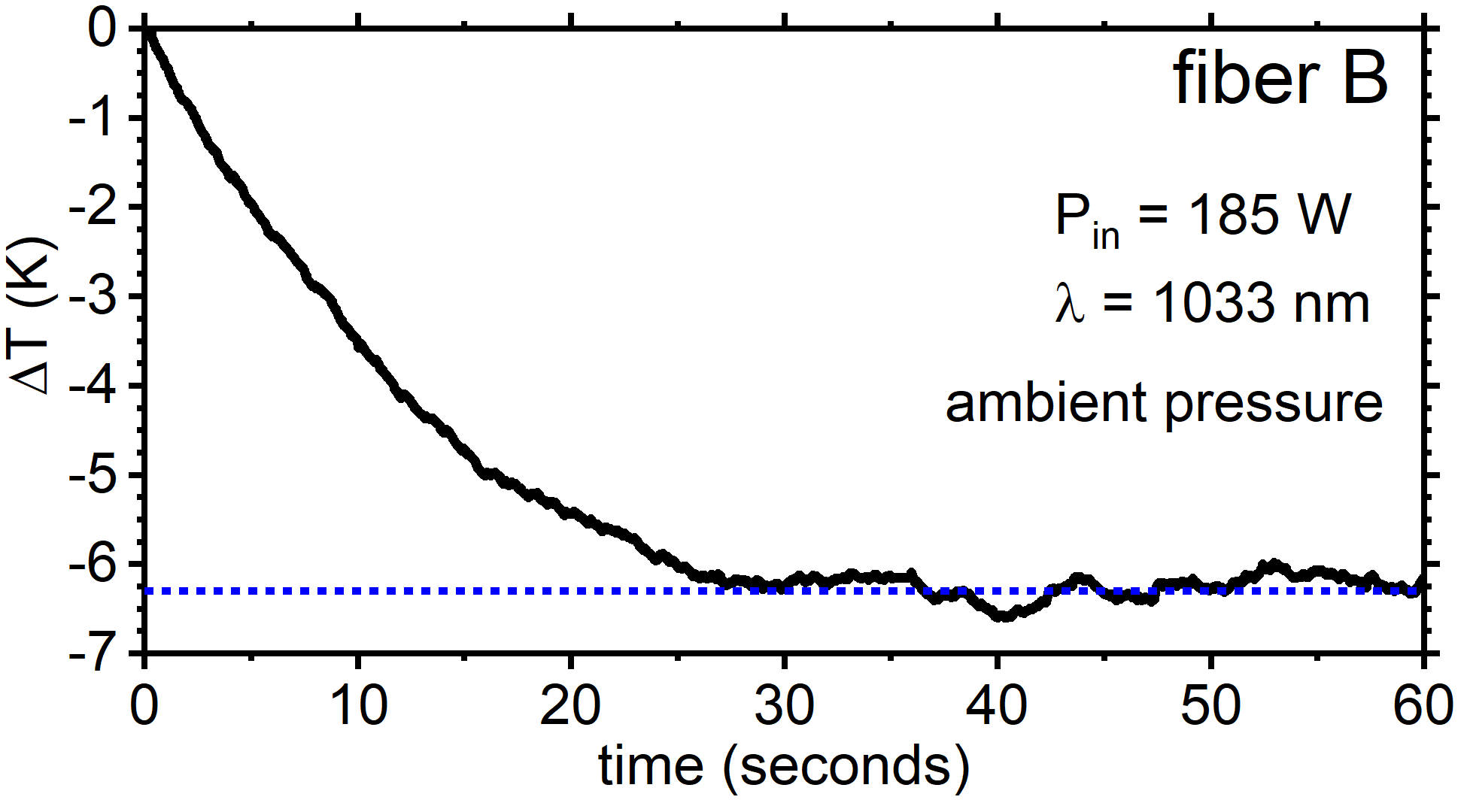}
\caption{Temporal cooling behavior of fiber B illuminated with 185\,W of 1033\,nm light under ambient pressure conditions. The dotted horizontal line is positioned at -6.3\,K to aid the eye.}
\label{fig:fiberB}
\end{figure}

\begin{figure}[t]
\centering
 \includegraphics[width=3.25 in]{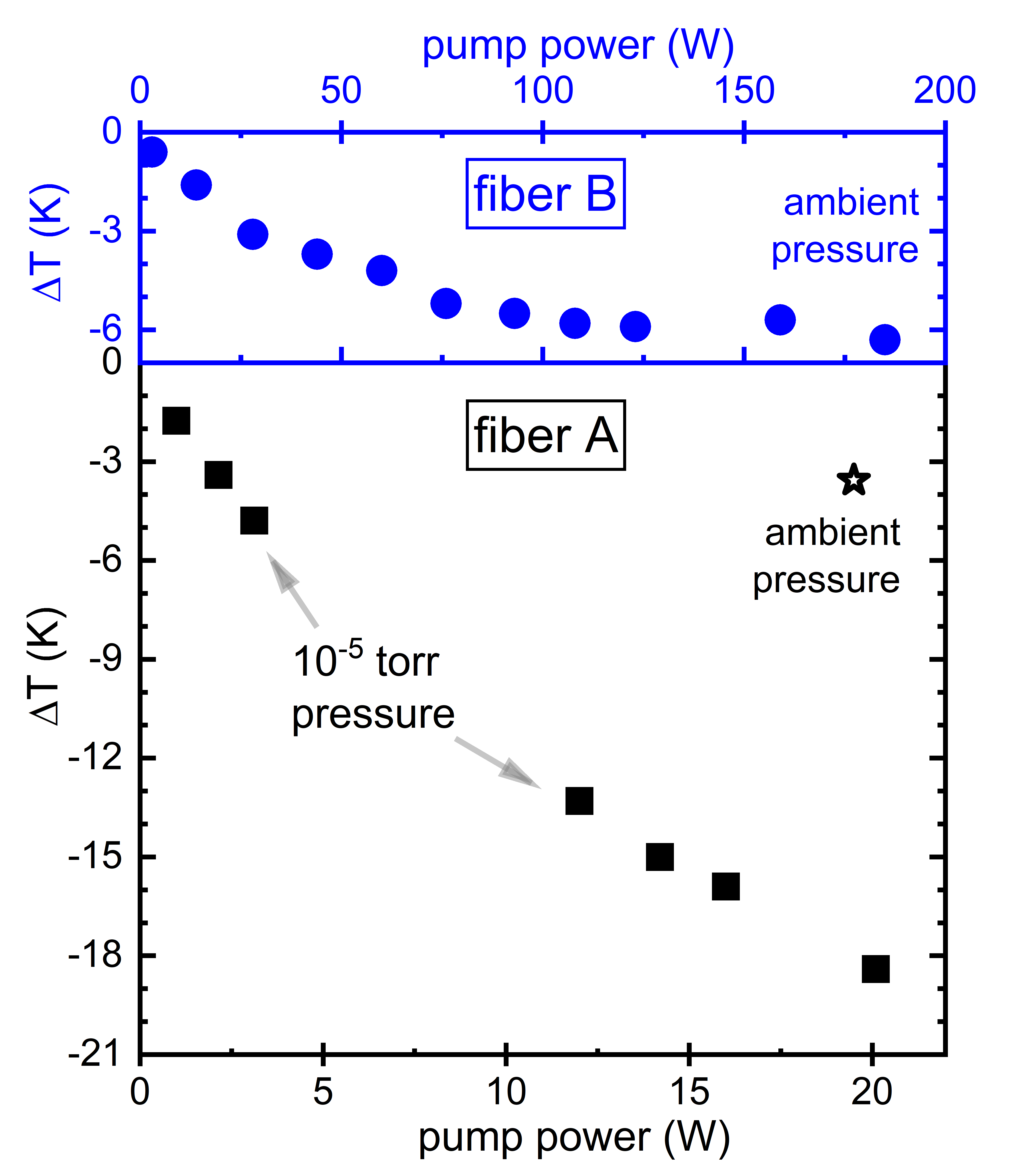}
\caption{Cooling for different pump powers for fiber A in-vacuum (black squares), fiber A in-air (open black star), and fiber B in-air (blue circles).}
\label{fig:alldata}
\end{figure}

For fiber B, experiments were conducted under ambient pressure conditions. Cooling by 6.3\,K from room temperature was observed for 185\,W of a 1033\,nm pump (Fig.\,\ref{fig:fiberB}). In Fig.~\ref{fig:alldata}, we plot the cooling as a function of pump power from all experimental trials. This illustrates well that we may reasonably expect cooling by more than 20\,K below room temperature in-vacuum with larger pump powers. The measurements on fiber A and fiber B were made at separate locations, and so it was not possible to use the vacuum cube configuration and the 185\,W amplifier simultaneously. This gap will be bridged in our future work. Giving consideration to Figs.~\ref{fig:cool} and~\ref{fig:alldata}, for the current fibers we anticipate cooling in-vacuum with higher pump powers to yield cooling to about 30\,K below room temperature.

In summary, for the first time, to the best of our knowledge, optically cooling silica in-vacuum to more than 18\,K from room temperature has been achieved. Compared to our previous work, increasing the Yb$^{3+}$ concentration and significantly reducing the thermal load of the passive cladding increased the cooling achieved by a factor of three. These results suggest that these fibers may serve as a platform for a desirable radiation-balanced laser. 

\smallskip
\smallskip

This material is based upon work supported by the Air Force Office of Scientific Research under award number FA9550-16-1-0362 titled Multidisciplinary Approaches to radiation-balanced Lasers (MARBLE).\\


\begin{thebibliography}{10}
\newcommand{\enquote}[1]{``#1''}

\bibitem{raman1928}
K.~Ramanm~C.V., Krishnan, \enquote{A new type of secondary radiation,}
  {{Nature}} \textbf{121}, 501--502 (1928).

\bibitem{pringsheim1929zwei}
P.~Pringsheim, \enquote{Zwei bemerkungen {\"u}ber den unterschied von
  lumineszenz-und temperaturstrahlung,} {{Zeitschrift
  f{\"u}r Physik}} \textbf{57}, 739--746 (1929).

\bibitem{epstein1995observation}
R.~I. Epstein, M.~I. Buchwald, B.~C. Edwards, T.~R. Gosnell, and C.~E. Mungan,
  \enquote{Observation of laser-induced fluorescent cooling of a solid,}
  {{Nature}} \textbf{377}, 500--503 (1995).

\bibitem{seletskiy2010laser}
D.~V. Seletskiy, S.~D. Melgaard, S.~Bigotta, A.~Di~Lieto, M.~Tonelli, and
  M.~Sheik-Bahae, \enquote{Laser cooling of solids to cryogenic temperatures,}
  {{Nature Photonics}} \textbf{4}, 161--164 (2010).

\bibitem{seletskiy2016laser}
D.~V. Seletskiy, R.~Epstein, and M.~Sheik-Bahae, \enquote{Laser cooling in
  solids: advances and prospects,} {{Reports on Progress
  in Physics}} \textbf{79}, 096401 (2016).

\bibitem{nemova2010laser}
G.~Nemova and R.~Kashyap, \enquote{Laser cooling of solids,}
  {{Reports on Progress in Physics}} \textbf{73}, 086501
  (2010).

\bibitem{hoyt2000observation}
C.~Hoyt, M.~Sheik-Bahae, R.~Epstein, B.~Edwards, and J.~Anderson,
  \enquote{Observation of anti-stokes fluorescence cooling in thulium-doped
  glass,} {{Physical Review Letters}} \textbf{85}, 3600
  (2000).

\bibitem{melgaard2016solid}
S.~D. Melgaard, A.~R. Albrecht, M.~P. Hehlen, and M.~Sheik-Bahae,
  \enquote{Solid-state optical refrigeration to sub-100 kelvin regime,}
  {{Scientific reports}} \textbf{6}, 1--6 (2016).

\bibitem{mobini2019laser}
E.~Mobini, S.~Rostami, M.~Peysokhan, A.~Albrecht, S.~Kuhn, S.~Hein, C.~Hupel,
  J.~Nold, N.~Haarlammert, T.~Schreiber \emph{et~al.}, \enquote{Laser cooling
  of silica glass,} {{arXiv preprint arXiv:1910.10609}}
  (2019).

\bibitem{mobini2020laser}
E.~Mobini, S.~Rostami, M.~Peysokhan, A.~Albrecht, S.~Kuhn, S.~Hein, C.~Hupel,
  J.~Nold, N.~Haarlammert, T.~Schreiber \emph{et~al.}, \enquote{Laser cooling
  of ytterbium-doped silica glass,} {{Communications
  Physics}} \textbf{3}, 1--6 (2020).

\bibitem{Knall:2020}
J.~Knall, M.~Engholm, J.~Ballato, P.~D. Dragic, N.~Yu, and M.~J.~F. Digonnet,
  \enquote{Experimental comparison of silica fibers for laser cooling,}
  {{Opt. Lett.}} \textbf{45}, 4020--4023 (2020).

\bibitem{10.1117/12.2510889}
J.~M. Knall, A.~Arora, P.~D. Dragic, J.~Ballato, M.~Cavillon, T.~Hawkins,
  S.~Jiang, T.~Luo, M.~Bernier, and M.~Digonnet, \enquote{{Experimental
  investigations of spectroscopy and anti-Stokes fluorescence cooling in
  Yb-doped silicate fibers},} in \emph{Photonic Heat Engines: Science and
  Applications,}  vol. 10936 D.~V. Seletskiy, R.~I. Epstein, and
  M.~Sheik-Bahae, eds., International Society for Optics and Photonics (SPIE,
  2019), pp. 40 -- 49.

\bibitem{PhysRevApplied.11.014066}
E.~Mobini, M.~Peysokhan, B.~Abaie, M.~P. Hehlen, and A.~Mafi,
  \enquote{Spectroscopic investigation of $\mathrm{Yb}$-doped silica glass for
  solid-state optical refrigeration,} {{Phys. Rev.
  Applied}} \textbf{11}, 014066 (2019).

\bibitem{8426483}
E.~{Mobini}, M.~{Peysokhan}, B.~{Abaie}, and A.~{Mafi}, \enquote{Investigation
  of solid state laser cooling in ytterbium-doped silica fibers,} in \emph{2018
  Conference on Lasers and Electro-Optics (CLEO),}  (2018), pp. 1--2.

\bibitem{10.1117/12.2545233}
E.~Mobini, S.~Rostami, M.~Peysokhan, A.~R. Albrecht, S.~Kuhn, S.~Hein,
  C.~Hupel, J.~Nold, N.~Haarlammert, T.~Schreiber, R.~Eberhardt,
  A.~Tünnermann, M.~Sheik-Bahae, and A.~Mafi, \enquote{{Observation of
  anti-Stokes fluorescence cooling of ytterbium-doped silica glass (Conference
  Presentation)},} in \emph{Photonic Heat Engines: Science and Applications
  II,}  vol. 11298 D.~V. Seletskiy, R.~I. Epstein, and M.~Sheik-Bahae, eds.,
  International Society for Optics and Photonics (SPIE, 2020).

\bibitem{10.1117/12.2548506}
J.~M. Knall, P.-B. Vigneron, M.~Engholm, P.~D. Dragic, N.~Yu, J.~Ballato,
  M.~Bernier, and M.~Digonnet, \enquote{{Experimental observation of cooling in
  Yb-doped silica fibers},} in \emph{Photonic Heat Engines: Science and
  Applications II,}  vol. 11298 D.~V. Seletskiy, R.~I. Epstein, and
  M.~Sheik-Bahae, eds., International Society for Optics and Photonics (SPIE,
  2020), pp. 48 -- 55.

\bibitem{Knall:20}
J.~Knall, P.-B. Vigneron, M.~Engholm, P.~D. Dragic, N.~Yu, J.~Ballato,
  M.~Bernier, and M.~J.~F. Digonnet, \enquote{Laser cooling in a silica optical
  fiber at atmospheric pressure,} {{Opt. Lett.}}
  \textbf{45}, 1092--1095 (2020).

\bibitem{Knall_20_comp}
J.~Knall, M.~Engholm, J.~Ballato, P.~D. Dragic, N.~Yu, and M.~J.~F. Digonnet,
  \enquote{Experimental comparison of silica fibers for laser cooling,}
  {{Opt. Lett.}} \textbf{45}, 4020--4023 (2020).

\bibitem{Peysokhan:2021}
M.~Peysokhan, S.~Rostami, E.~Mobini, A.~R. Albrecht, S.~Kuhn, S.~Hein,
  C.~Hupel, J.~Nold, N.~Haarlammert, T.~Schreiber, R.~Eberhardt, A.~Flores,
  A.~Tünnermann, M.~Sheik-Bahae, and A.~Mafi, \enquote{Implementation of
  laser-induced anti-stokes fluorescence power cooling of ytterbium-doped
  silica glass,} {{ACS Omega}} \textbf{6}, 8376--8381
  (2021). PMID: 33817498.

\bibitem{richardson2010review}
D.~Richardson, J.~Nilsson, and W.~Clarkson, \enquote{High power fiber lasers:
  current status and future perspectives [invited],} {{J.
  Opt. Soc. Am. B}} \textbf{27}, B63--B92 (2010).

\bibitem{brown2001}
D.~C. Brown and H.~J. Hoffman, \enquote{Thermal, stress, and thermo-optic
  effects in high average power double-clad silica fiber lasers,}
  {{IEEE J. Quantum Electron}} \textbf{37}, 207--217
  (2001).

\bibitem{zenteno1993}
L.~Zenteno, \enquote{High-power double-clad fiber lasers,}
  {{J. Lightwave Technol}} \textbf{11}, 1435--1446 (1993).

\bibitem{ward2012}
B.~Ward, C.~Robin, and I.~Dajani, \enquote{Origin of thermal modal
  instabilities in large mode area fiber amplifiers,}
  {{Opt. Express}} \textbf{20}, 11407--11422 (2012).

\bibitem{dawson2008}
J.~W. Dawson, M.~J. Messerly, R.~J. Beach, M.~Y. Shverdin, E.~A. Stappaerts,
  A.~K. Sridharan, P.~H. Pax, J.~E. Heebner, C.~W. Siders, and C.~Barty,
  \enquote{Analysis of the scalability of diffraction-limited fiber lasers and
  amplifiers to high average power,} {{Opt. Express}}
  \textbf{16}, 13240--13266 (2008).

\bibitem{peysokhan2020characterization}
M.~Peysokhan, E.~Mobini, A.~Allahverdi, B.~Abaie, and A.~Mafi,
  \enquote{Characterization of yb-doped zblan fiber as a platform for
  radiation-balanced lasers,} {{Photonics Research}}
  \textbf{8}, 202--210 (2020).

\bibitem{peysokhan2019measuring}
M.~Peysokhan, E.~Mobini, and A.~Mafi, \enquote{Measuring the anti-stokes
  cooling parameters of a yb-doped zblan fiber for radiation balancing,} in
  \emph{Sixth International Workshop on Specialty Optical Fibers and Their
  Applications (WSOF 2019),}  vol. 11206 (2019), pp. 112061Q--1.

\bibitem{bowman1999}
S.~Bowman, \enquote{Lasers without internal heat generation,}
  {{IEEE Journal of Quantum Electronics}} \textbf{35},
  115--122 (1999).

\bibitem{bowman2010}
S.~R. Bowman, S.~P. O’Connor, S.~Biswal, N.~J. Condon, and A.~Rosenberg,
  \enquote{Minimizing heat generation in solid-state lasers,}
  {{IEEE J. Quantum Electron}} \textbf{46}, 1076--1085
  (2010).

\bibitem{bowman2016}
S.~Bowman, \enquote{Low quantum defect laser performance,}
  {{Opt. Eng.}} \textbf{56}, 011104 (2016).

\bibitem{knall2021radiationbalanced}
J.~Knall, M.~Engholm, T.~Boilard, M.~Bernier, P.-B. Vigneron, N.~Yu, P.~Dragic,
  J.~Ballato, and M.~Digonnet, \enquote{Radiation-balanced silica fiber laser,}
  {{Optica}} \textbf{8}, 830--833 (2021).

\bibitem{knallRBL}
J.~M. Knall, M.~Engholm, T.~Boilard, M.~Bernier, and M.~J.~F. Digonnet,
  \enquote{Radiation-balanced silica fiber amplifier,}
  {{Phys. Rev. Lett.}} \textbf{127}, 013903 (2021).

\bibitem{kuhn2019}
S.~Kuhn, S.~Hein, C.~Hupel, J.~Nold, F.~Stutzki, N.~Haarlammert, T.~Schreiber,
  R.~Eberhardt, and A.~Tünnermann, \enquote{High-power fiber laser materials:
  influence of fabrication methods and codopants on optical properties,} in
  \emph{Optical Components and Materias XVI,}  vol. 10914 S.~Jiang and M.~J.
  Digonnet, eds., International Society for Optics and Photonics (SPIE, 2019),
  pp. 15 -- 27.

\bibitem{beier2017}
F.~Beier, C.~Hupel, S.~Kuhn, S.~Hein, J.~Nold, F.~Proske, B.~Sattler, A.~Liem,
  C.~Jauregui, J.~Limpert, N.~Haarlammert, T.~Schreiber, R.~Eberhardt, and
  A.~Tünnermann, {{Opt. Express}} \textbf{25},
  14892--14899 (2017).

\bibitem{beier2018}
F.~Beier, F.~Möller, B.~Sattler, J.~Nold, A.~Liem, C.~Hupel, S.~Kuhn, S.~Hein,
  N.~Haarlammert, T.~Schreiber, R.~Eberhardt, , and A.~Tünnermann,
  \enquote{Experimental investigations on the tmi thresholds of low-na yb-doped
  single-mode fibers,} {{Opt. Lett.}} \textbf{43},
  1291--1294 (2018).

\bibitem{imangholi2006}
B.~Imangholi, M.~P. Hasselbeck, D.~A. Bender, C.~Wang, M.~Sheik-Bahae, R.~I.
  Epstein, and S.~Kurtz, \enquote{Differential luminescence thermometry in
  semiconductor laser cooling,} in \emph{Physics and Simulation of
  Optoelectronic Devices XIV,}  vol. 6115 (International Society for Optics and
  Photonics, 2006), p. 61151C.

\end{thebibliography}
\end{document}